# Design, Construction, and Use of a Single Board Computer Beowulf Cluster: Application of the Small-Footprint, Low-Cost, InSignal 5420 Octa Board

James J. Cusick, William Miller, Nicholas Laurita, Tasha Pitt
{James.Cusick; William.Miller; Nicholas.Lauita; Tasha.Pitt}@wolterskluwer.com
Wolters Kluwer, New York, NY

*Abstract*—In recent years development in the area of Single Board Computing has been advancing rapidly. At Wolters Kluwer's Corporate Legal Services Division a prototyping effort was undertaken to establish the utility of such devices for practical and general computing needs. This paper presents the background of this work, the design and construction of a 64 core 96 GHz cluster, and their possibility of yielding approximately 400 GFLOPs from a set of small footprint InSignal boards created for just over $2,300. Additionally this paper discusses the software environment on the cluster, the use of a standard Beowulf library and its operation, as well as other software application uses including Elastic Search and ownCloud. Finally, consideration will be given to the future use of such technologies in a business setting in order to introduce new Open Source technologies, reduce computing costs, and improve Time to Market.

## I. Introduction

For any company running a computing infrastructure things are always changing. For people who have been around a while we remember shifts from centralized computers to distributed computers to networked client/server architectures to Intranet and Internet applications and finally to mobile architectures (and others too). Our interest around Single Board Computers (SBCs) developed in 2013 with the Raspberry Pi which became a phenomenon. For $35 we were able to buy a fully functioning Linux server and, for example, create a cloud repository. This got us thinking about what else we might be able to do with such technologies at such an attractive price point.

But first a word about whom we are. Wolters Kluwer (WK) is a Netherlands-based international publisher and digital information services provider with operations around the world. Wolters Kluwer is organized into Business Units which then control operating companies. The experience documented here focuses on work done for the New York-based Corporate Legal Services (CLS) Division which manages five units including CT Corporation (CT). The systems operated by CT include public-facing Web-based applications and internally used ERP (Enterprise Resource Planning) systems. Major vendors manage network services and hosting for our computing environments. The CT IT team is responsible for the development and operations of these systems from an end customer standpoint. The R&D work reflected here is hoped to benefit the development of these systems in the long run by reducing costs, improving flexibility, and reducing Time to Market through more rapid technology adoption.

This paper first provides some background on SBCs and what our goals were in pursuing their application within a structured Proof of Concept (POC) project. We then review the design of the cluster from the specifications of the InSignal board up to the software environment layers. We will present the costs of the components required to build the cluster, the stumbling blocks we ran into and their solutions, as well as some things we were unable to solve. The paper also covers the configuration steps needed to establish the Beowulf clustering software and the verification tests we ran. Finally we discuss the future plans for the cluster and reflect on what these types of devices might mean for computing architectures in the future.

## II. Background

For CT there is a continuous demand for computing environments with both cost efficiency and flexibility. Currently CT spends large sums annually on computing resources which are sometimes underutilized and at other times oversubscribed. There is a constant need for rapid realignment of resources but the design of the environments does not always easily allow for this. In pursuing this POC CT IT planned to gain critical knowledge of emerging compact, high performance, low cost, rapidly evolving SBCs and the open source software solutions they can support.

There have been recent developments around Single Board Computing (SBC) or microcomputers (aka, System on a Chip – SoC). These extremely small form factor computers provide significant computing capabilities for very low cost. Typical devices range from $50 to $200 and are provided on a circuit board of about the size of a 3x5 index card or less. These devices run at extremely low power (typically 5v), provide multiple interface options, and normally run a variant of Linux. Processor speeds vary but are most are below 1 GHz. RAM is often between 512 MB and 1 GB. Today these devices are used widely by hobbyists, educators, and innovative IT groups. There are some new models that exceed these low performance limits and these will be the focus of this POC.

During a recent conference held by CUNY in Manhattan there was a showcase of projects around the use of Raspberry Pi's [1]. This demonstration peaked our interest in these technologies and what they might be capable of in our business. In our research there were dozens of SBCs available. Some of the most popular at the time included:

- **Raspberry Pi**: Probably the most popular device which was developed in the UK at Cambridge for the purposes of educating students on computing. To date over 4,000,000 Pi's have been shipped. The Raspberry Pi has a 700 MHz CPU which can be overclocked to 1 GHz and 512 MB RAM. It supports Ethernet, 2 USB ports, HDMI, audio, video, and power [2]. The cost of a model B is $35 but it does require a few other components which in our case brought the cost to about $65 before tax and shipping. We will discuss an early prototype effort using the Pi below.
- **BeagleBoard**: This is another famous hobbyist-friendly, single board computer. It costs US $149 and has an open source design. The system is USB powered and runs a Texas Instruments OMAP 3530 system-on-a-chip (SoC), which has a 600MHz ARM Cortex A8 processor.
- **PandaBoard**: For just over $170 this mobility-friendly single-board computer based on the TI OMAP4430 SoC, includes HDMI, 10/100 Ethernet, Wi-Fi, Bluetooth, and multiple USB connectors.
- **Via APC**: Via announced a US $49 computer runs Google's Android platform. It includes 512MB of DDR3 RAM, 2GB of flash storage and supports HDMI, D-sub/VGA, four USB 2.0 ports, audio jacks, Ethernet, and a microSD slot. The APC will use VIA's own 800MHz processor and run a version of Android 2.3 at launch. It measures 170x85mm.

III. ORIGINAL PI PROTOTYPE

To begin exploring the SBC concept a Raspberry Pi prototype environment was constructed. The purpose of the prototype was to explore the utility of the Pi, implement a working project, and assess the use of the device for CT development purposes. After researching the device the hardware was ordered via Amazon.com. The project selected to prove out the environment was to set up a personal/shared cloud using ownCloud which is a freeware/open source environment. There are an unlimited number of projects that could be selected for further work. By searching the Internet for "Raspberry Pi projects" many interesting ideas will be provided. To create a working system, in addition to the Raspberry Pi Model B, the following items were ordered:

1. A power supply (5v 1500ma USB micro power supply)
2. A case (there are many types available)
3. An 8 GB SD card (with multiple versions of Linux)
4. A 1 TB external hard drive
5. A VGA monitor and an HDMI to VGA converter cable
6. A USB keyboard and mouse
7. A USB hub

The Pi and parts itself cost $65, the cost of the entire system excluding monitor and keyboard was $200 including shipping and taxes. The platform is shown in Figure 1 below. Setting up the equipment took only a few minutes. After installing Raspbian (the version of Debian Linux for the Pi) and installing a number of other packages including a C compiler, Python, and other programming tools the core environment was ready. To install and run ownCloud Apache is required and needs configuring. Once the settings were correct one could access the cloud repository via multiple web browsers from multiple laptops and tablets on the LAN. The prototype was not set up for public access but this can be done simply by configuring an external IP address.

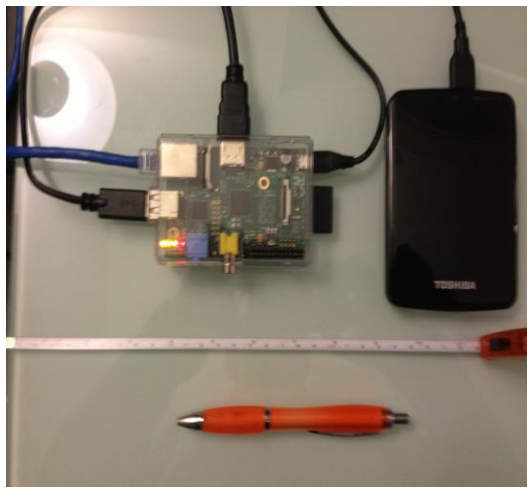

*Figure 1 – Raspberry Pi for OwnCloud POC*

IV. MOVING TOWARDS THE CLUSTER POC

*A. Possibilities for the Pi at CT*

Initially the Pi was felt to have good possibilities for supporting some CT development needs. We could certainly use a Pi for some development work here or there. There is no question that at $65 a Linux device could prove handy for some purposes. An initial idea was to use it for creating a Linux based Oracle database test environment but that might overwhelm a single device or may not even be compatible with the device (later, in speaking with our Oracle technical support team they did provide information on how they have created some Pi based database prototypes but with alternate ARM compliant applications and no their standard enterprise

software offerings [3]). However, various people have published work on creating Raspberry Pi clusters with very high throughput. Researchers in the UK created a cluster of Raspberry's to form a supercomputer which cost roughly $3,000[4]. This consisted of 64 Pi's with 1 TB of RAM and performed 1.8 GFLOPS. A similar device was created in the US with a 32 node Raspberry cluster for about $1,900 and competes with some supercomputer performance metrics [5]. This device and others have employed the standard Beowulf cluster libraries to build parallel computing environments of varying sizes form SBCs including Kiepert's [5]. Thus, the Pi has been proven a useful machine for application as a general purpose Linux development box, web server, file server, or clustered host for a database. However, we did not think the Pi was the right device for a larger scale project due to its limited performance characteristics so we started looking at other products.

### B. The Parallella SBC

After further research a newly released SBC came up. The Parallella from Adapteva is an extremely interesting device utilizing massively parallel chip architecture. The Adapteva Epiphiny chip has a dense mesh of matrixed cores providing high scalability. Their product called the Parallella [6] offers a very powerful, low cost, small form factor parallel computer. The board provides 16 cores with 1 GB of RAM. The processors run at 800 GHz providing 12.8 GHz computing for about $150. Parallella also offers a 4 board cluster of 16 cores providing 64 cores originally priced at $575 running at about 100 GFLOPS. Thus, for a price far below a low-end PC you get computing power rivaling standard offerings. Best of all, the basic boards are only 3" x 2" so you can fit one in your pocket. We planned to us the Parallella for this POC however the company ran into both supply chain and engineering issues with the board (specifically heat displacement problems) in 2014 during its initial rollout. As a result we moved on to another manufacturer called InSignal although the Parallella is now available.

### C. The Octa Board

Once we determined that we needed to move to another product there were only a few next tier providers in terms of equivalent device capabilities in early 2014. One of them was InSignal with their Arndale 5420 Octa Board [7]. After comparing with other similar 8 core devices which represented the most powerful devices at the time we jumped in and bought 8 of the boards to allow us to construct a 64 core cluster somewhat comparable to the Parallella cluster.

The Samsung Exynos 5420 Octa Board (see Figure 2) is a system-on-chip (SoC) based on 32-bit RISC processor for smartphones, tablets, laptops, and desktop PCs. The Exynos 5420 adopts a *big.Little* architecture using the Cortex-A15 core (quad) and targets 1.8GHz speed. It also incorporates the Cortex-A7 core (quad) which enables energy efficient computing for less intensive tasks running at 1.2 GHz. The Exynos 5420 provides 14.9GB/s memory bandwidth for heavy traffic operations such as 1080p video en/decoding, 3D graphics display and high resolution image signal processing with WQXGA Display. The application processor supports dynamic virtual address mapping aiding software engineers to fully utilize the memory resources easily. The board layout appears in Figure 2 below and the actual size of the board is demonstrated next to a pen in Figure 3 below.

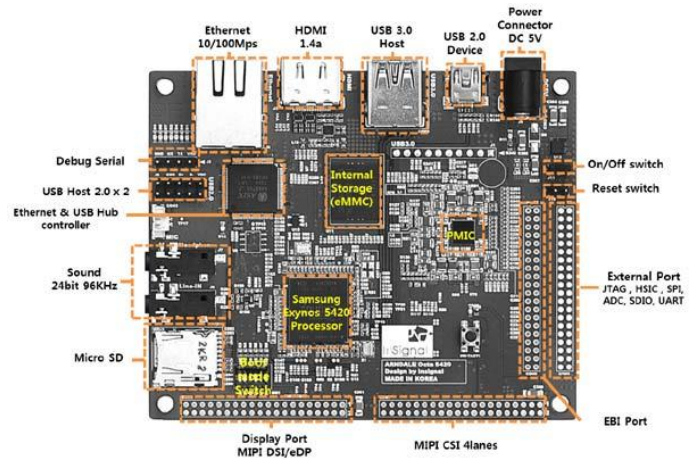

*Figure 2 – The Octa Board Layout*

### D. Octa Board Core Specifications

- ARM Cortex™-A15 Quad 1.8GHz
- ARM Cortex™-A7 Quad 1.3GHz
- Memory LPDDR3e (14.9GB/s bandwidth) 3GB
- 32KB(Inst)/32KB(Data) L1 Cache & 2MB L2 Cache
- 1TB physical addressing
- Dimensions: 104mm length X 85mm width

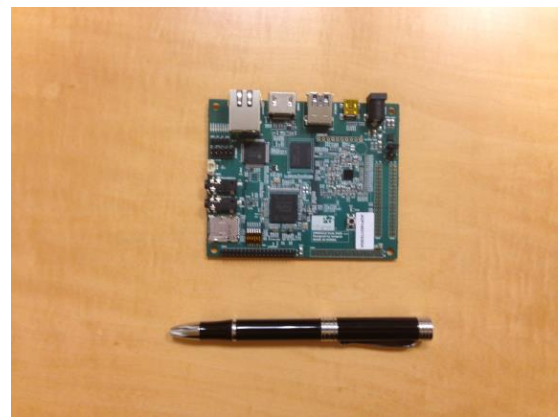

*Figure 3 – The Octa Board Relative Size*

## V. APPLICATIONS OF THE OCTA BOARD

### A. The POC Objective

As mentioned in the introduction, CT has significant computing needs and at the moment they are not always met in as flexible a manner as desired and often at a higher cost than desired. The general problem CT development encounters from the computing environment standpoint include:

1. Cost
2. Inadequate flexibility and complex scheduling needs
3. Limited developer administration rights
4. Complexity of environment architectures
5. Requirements for dedicated test data environments

There are certainly other issues we face with our computing environments. However, this POC focused primarily on several items as follows:

- Exploring the capabilities of SBC devices, specifically the Octa Board, so as to be able to utilize these machines for rapid R&D work in the future and do so for less than $2,000. Develop the in-house knowledge to manipulate and extend the use of these individual devices for R&D purposes.
- Construct an SBC cluster to achieve high performance computing capabilities using the Octa Board so as to support larger computing needs than only server at a time environments might not allow in development. This cluster became known as the "Octa-Cluster".
- Attempt to provide or replace an Oracle development database environment to realize a significant cost savings. An SBC cluster environment could provide sufficient processing power at orders of magnitude less cost to carry out this work.
- Prove that if it is possible to provide a suitable and versatile development environment for $2K it may be possible to create additional development environments thereby creating entirely new development options. By scaling out in this manner, access and flexibility to development (or even QA) environments can be greatly enhanced at very limited cost.
- Prove that the SBC based environment can easily be cloned. Thus for very limited cost multiple development environments can be created and each one can be opened up to developers to manipulate with administrative rights as they can be replaced or refreshed very easily.
- Create the foundations for future research into alternative internally hosted, low cost, small footprint, highly adaptable computing environments to enable new development within CT.

## VI. THE CLUSTER ARCHITECTURE

### A. Generic Beowulf Cluster Approach

In thinking about how to cluster the Octa Boards we quickly hit on the idea of creating a Beowulf cluster. The concept and approach to a Beowulf cluster has been well established starting in the early 1990s based on work done at NASA Goddard by Thomas Sterling and Donald Becker [8]. The definition of a Beowulf cluster follows:

*"A Beowulf cluster is a group of what are normally identical, commercially available computers, which are running a Free and Open Source Software (FOSS), Unix-like operating system, such as BSD, GNU/Linux, or Solaris. They are networked into a small TCP/IP LAN, and have libraries and programs installed which allow processing to be shared among them."[9]*

A true Beowulf is a cluster of computers interconnected with a network with the following characteristics [9]:

1. The nodes are dedicated to the Beowulf cluster.
2. The network on which the nodes reside are dedicated to the Beowulf cluster.
3. The nodes are Mass Market Commercial-Off-The-Shelf (M2COTS) computers.
4. The network is also a COTS entity.
5. The nodes all run open source software.
6. The resulting cluster is used for High Performance Computing (HPC).

### B. The Octa-Cluster Design

For the CT SBC Cluster (aka, "Octa-Cluster") we followed the generic Beowulf cluster architecture closely but not perfectly. The cluster architecture is shown in Figure 4 below. The master node was provided any one of the Octa Board nodes where an application was instantiated under the clustering software. MPI clustering software (http://www.open-mpi.org/) is installed on each node to allow for parallel processing across the full platform. An Intel based HP Tower running Ubuntu server provided the external mass storage array access. A total of 4 TB of external storage was attached to the master node via USB 3.0 and made available to each Octa Board nodes mounted via NFS (Network File System). A 16 port 10/100 Mbps unmanaged switch bridges the 8 Arndale 5420 nodes and the host server and also provides access to the broader network which does violate the pure Beowulf architecture of running on an isolated network. We felt this could be corrected in future revisions to the implementation by adding a secondary NIC card to the cluster host.

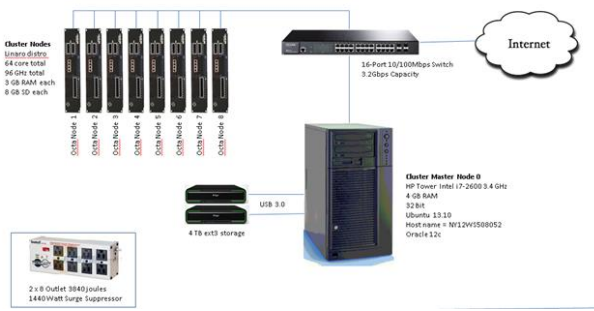

*Figure 4 – The Octa-Cluster Architecture*

### C. Steps for Creating the Cluster Environment

1. Develop project concept

2. Document the plan
3. Submit plan with costs for approval
4. Equipment research
5. Equipment orders
6. Receive all equipment and setup
7. Install Ubuntu on master PC
8. Configure 4 TB storage for master PC
9. Set up initial Octa board and configure
10. Setup and install all 8 Octa boards with OS and Linux configuration
11. Networking configuration and switch set up
12. Configure NFS
13. Install application synchronization software library (MPI)
14. Baseline Octa-cluster performance via parallel processing benchmarks
15. Install additional software: Apache, C++, Ruby, MongoDB, MySQL to explore other technology capabilities of platform including ownCloud
16. Install and configure Elastic Search for additional parallel processing demonstration application

## VII. INSTALLATION EXPERIENCE

### A. Some Preparatory Discussions

In early meetings with the Oracle embedded systems support team to review our plan [3], Oracle noted they had done a Raspberry POC recently. They also noted many customers doing embedded systems work were working with SBC devices, however, they had not seen any one doing a cluster using an SBC platform as we were proposing. They said that as long as there was an Oracle build for the ARM processor (InSignal's 5420 is powered by a Samsung Exynos ARM processor) then this should work but we might need to be flexible and look at a compact Oracle deployment and not the Enterprise version. One possibility was to deploy the Oracle application on the master node/server and do the processing across the cluster. This could be another working scenario but would violate the principles of the Beowulf clustering we were working with.

### B. Getting the Boards Running

Once the Octa Boards were delivered we began setting them up as seen in Figure 5 below. Initially we connected to the Octa Board via the HDMI port using a VGA converter cable which we had previously used successfully with the Pi. This did not allow for a successful startup so we switched to a DB9 serial cable and began setting up the *ttyS0* port on the Ubuntu server for communications to initial test board. After some challenges with settings, eventually we got a successful connection to the board; however, we did not get a login prompt from the board or its expected *uboot* interface. At least none was detected.

Eventually we decided to give one card each to a couple of the team members to take home and experiment with since time in the office is typically hard to come by. They started to look into the issues with the board. In one case the only monitor available at one of the home locations was an HDMI TV. A connection to one of the boards via the HDMI port on the TV resulted in a basic screen image. Also, a successful connection using a USB to 9 pin converter to the DB9 connector on the 5420 was received and a command prompt over *Putty* was established. We then purchased a Samsung HDMI monitor for the lab in the office and plugged it into an eMMC (Embedded Multi Media Card) configured board. We were able to boot the board to a default Android splash screen.

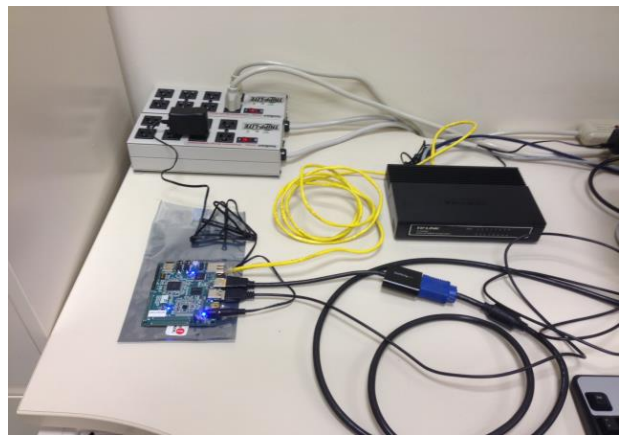

*Figure 5 – Initial Installation of the Octa Board*

Booting the board from eMMC launched a preinstalled version of Android. We attached a USB hub with keyboard and mouse and logged into the environment for the first time and were able to execute some basic commands via the GUI. However, the Android version was very unstable. It crashed within a few minutes in most cases. However, we tested each board to make sure each of them could all boot up into Android and all were in fact successful so we were confident we did not have any DOA boards.

We then switched our efforts to converting the boards to run Linux which was our target environment. We purchased a USB microSD-to-SD adapter and began creating Linux images for booting experimentation. This required some trial and error and a few rounds of repartitioning and recreating images for a working version of Ubuntu on an initial microSD . We had copied down the Linaro (http://www.linaro.org/) AMD image of Linux for the Arndale Octa Board and installed it on the microSD. Once we had the layout correct we then planned to boot the board from the microSD with Linaro Linux as opposed to Android. This turned out to be the right approach to getting the board running in our environment compared with any other method we attempted.

The relabeling of the microSD card back to "*/boot*" was done with a utility called *UNetbootin* to create bootable images for Linaro. This succeeded but the results with the board were not clear at first. Actually the same poor results with the serial connection where no commands were received was observed at first, however, with the bootable Linaro microSD the RJ45 Ethernet jack did light up (but was not stable). With the earlier pre-built version the port did not light up at all.

In working on booting from the microSD, we downloaded the latest Linaro distribution, flipped the appropriate dip switch settings on the board from the default settings (all 6 switches off) to having only the #3 switch up as documented but booting was still not successful. We attempted to try other switch settings but we remained unsuccessful. After re-creating the boot image on the microSD several times to try to make sure it was correct we were still not successful. We eventually succeeded by in formatting the microSD and burning a bootable Linaro image. The application we used to create this image can be found at this location: http://win32diskimager.sourceforge.net/. Figure 6 shows the first successful login to the environment.

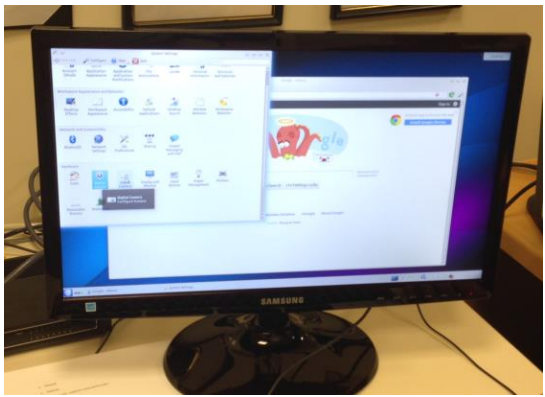

*Figure 6 – The first working GUI session on the Octa Board*

After the first board became operational and stable we began configuring the OS, adding root password, personal user accounts, *openssh* server, and tested remote login via *putty ssh*, created *sudo*, installed Apache, turned on *ftp* services, tested the web server, and installed the GUI of preference - KDE. Finally, we set up a basic web page so that the cluster had a presence on the Intranet.

At this point we began assembling the cluster. As seen in Figure 7 below, we built up the cluster one board at a time and wired the boards into the switch and then routed the power sources as appropriate. The completed cluster is shown in Figure 8 minus the host server and storage which are just adjacent to the cluster itself in the lab room. With the physical components in place and the base Operating System configured the POC turned to some of it's proving tasks.

### C.  Additional Problems Encountered

There were a number of issues we encountered in building out the Octa-Cluster. Not all of them have been solved at this time. Below are a few of these issues:

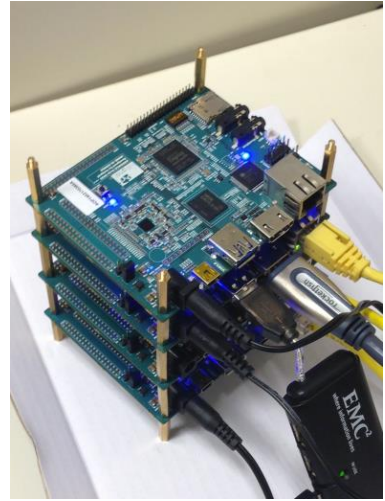

*Figure 7 – The Octa-Cluster under construction with 4 servers*

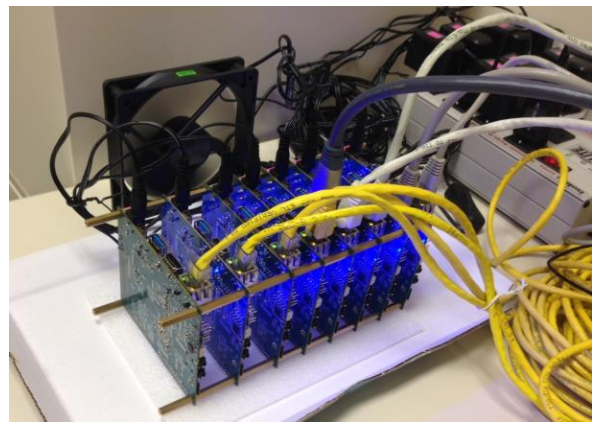

*Figure 8 – The Octa-Cluster in completed form with 5v power supplies, power conditioner, and fan in background*

- The monitor, acting as the console device, goes to sleep after hitting its idle time as expected. To get out of sleep mode, it should simply require some input whether movement of a mouse or a keyboard input. However, this does not happen consistently. The problem may be that the USB controller may be in sleep mode and therefore keyboard and mouse input goes undetected. Recovery sometimes requires rebooting the board. We can continue working on the board itself remotely.
- At varying times we have gotten a particular server (node) into such a state through installations of software and

experimental configurations that it was easier to reimage the microSD card with the OS then to try to debug the node environment. We have a very straightforward documented set of steps for this which only takes about an hour or two to complete this rebuild. This makes it easy to recover from those cases where our experimentations go awry.

- At one point during the configuration of the cluster we encountered some confusing network behaviors. Primarily these issues had to do with conflicting IP addresses. The solution for this was primarily to convert all nodes in the cluster to static IPs and also to insert these IPs in */etc/hostfile*. After some testing the cluster performed just fine.

### D. Parts Inventory and Costs

Below in Table 1 the parts and costs including shipping required to construct the Octa-Cluster are listed.

| Item | Qty | Item Price | Total with Shipping |
|---|---|---|---|
| InSignal Arndale Octa 5420s | 8 | $179.00 | $1,510.35 |
| 16GB microSD cards | 8 | $11.99 | $105.30 |
| DC 5V Power Supply | 8 | $7.97 | $63.76 |
| Surge protector – 14 outlet, 750 Watt | 2 | $124.16 | $248.32 |
| 12 port Switch | 1 | $29.74 | $42.35 |
| Cat 5e cables (6 in) | 8 | $0.52 | $5.77 |
| Storage: 3 TB | 1 | $109.99 | $109.99 |
| 2 PC USB fans | 2 | $27.98 | $55.96 |
| Board PCB standoff spacers | 1 | $13.11 | $13.11 |
| Board spacer nuts and bolts | 1 | $4.97 | $4.97 |
| DB9 serial cable | 1 | $6.41 | $14.41 |
| Dynex SD/MicroSD (USB Read/Write) | 1 | $11.96 | $11.96 |
| Samsung "20 LED HD HDMI Monitor | 1 | $141.53 | $141.53 |
| 6FT HDMI High Speed Cable | 1 | $32.65 | $32.65 |
| TOTAL | | | $2,360.43 |

Table 1 – Parts Inventory and Costs

### VIII. OPERATIONAL EXPERIENCE

#### A. Major Milestones of Operation

The first parts of the Octa Cluster arrived in early April 2014. By mid-June the cluster was operational. Since then the cluster has been running with no down time, through December 2014. During these six months we have installed the various software packages listed above and also carried out various tests and benchmarking. The two primary factors elongating the setup period were the confusion around the native HDMI requirement and the Operating System build approach especially getting the OS imaged properly on the microSD cards. These two issues probably added several weeks to the build out process. Today we can access the cluster locally from the console or GUI, remotely via telnet or Putty, and also across the VPN. The cluster runs 24x7 and we have approximately 8 developers and engineers accessing the environment at different times based on administrative needs or project requirements and activities.

#### B. Parallel Processing Test

Once the cluster was operational a standard benchmarking test was run to prove out the parallel computing function of the cluster. The benchmarking demo application *perftest* was program downloaded and installed [10]. After a few false starts the program ran but required some additional configuration and path setting changes to become usable. By creating a simple shell script the program eventually exercised all 64 cores in the cluster demonstrating successfully that all processors in the cluster were participating in a parallel execution of the same program. A sample of the test results are shown below:

```
mpiu@octa-node1:/root$ cd /mirror
mpiu@octa-node1:/mirror$ ./run_test.sh
Hello from processor 0 of 44 on server octa-node1
Hello from processor 1 of 44 on server octa-node1
Hello from processor 12 of 44 on server octa-node4
Hello from processor 13 of 44 on server octa-node4
Hello from processor 21 of 44 on server octa-node5
Hello from processor 2 of 44 on server octa-node2
Hello from processor 15 of 44 on server octa-node4
Hello from processor 22 of 44 on server octa-node5
Hello from processor 3 of 44 on server octa-node2
Hello from processor 17 of 44 on server octa-node4
Hello from processor 24 of 44 on server octa-node5
```

This output listing shows execution of commands across multiple processors and different node servers.

#### C. ownCloud Deployment

In order to test the multi-user access to the platform *ownCloud* was installed and configured. *ownCloud* [11] offers both a freeware and a paid usage version that allows for document management functionalities including individual user accounts, administration, document upload and download, document sharing among users, and document change notification. The environment was set up utilizing the NFS drive so as to allow for adequate storage per user as the Octa Board local storage is only 16 GB less the Operating System requirements. The performance of the application was acceptable both locally on the network and over the VPN. Multiple users were able to access the repository simultaneously with no observable performance degradation. The experience on the Octa Board was very similar to that on the prototype Pi environment. This software application is a good substitute for applications such as Microsoft's Sharepoint or other document sharing tools. It also provides a more secure document sharing solution than public cloud based solutions like Dropbox or Box as it resides inside your network. Being able to host this freeware software on a performant device which costs below $200 is attractive. Placing the software on the cluster provides some scalability and with proper hardening of the administration around the environment this could be a highly beneficial use of SBC technologies in general for business users.

### D. Elastic Search Test

Following this testing we can began working with Elastic Search to explore additional applications for the platform. The installation was straight forward; it installed smoothly when following Elastic Search's Linux instructions. CT had already been working on an independent and parallel project evaluating Elastic Search on a standalone Intel server. This provided good baseline metrics to compare performance. The approach was to index one of our primary tables having 2.3 million records. On an I7 Intel box with 8Gb of RAM this table was indexed within 4 hours. The initialization algorithm submitted a block of 100 records to Elastic Search for indexing until there were no more to submit.

On the Octa Cluster, this same approach was attempted by configuring one node to have a like comparison to the Intel baseline. The single Octa Board configuration failed with timeout errors after submitting the first block of 100. The block size was adjusted to larger and smaller ones. No matter the block size, failures were observed after the first submission with the larger block sizes and after a few submissions with the smaller ones.

To bring more power to the problem, Elastic Search was reconfigured into a cluster: first with 2 nodes, then 3, and finally 4. The alternative configurations were able to index at most 1900 records before observing failures. In all instances, Elastic Search climbed quickly to 100% CPU utilization and 98% RAM utilization. Once the 4 node cluster configuration had failures, we concluded that adding additional nodes was neither practical nor likely to solve the problem. The Elastic Search on the Octa-Cluster requires more research to evaluate the source of the performance bottleneck, but we speculate it is related to insufficient RAM resources.

### E. MongoDB Shakeout

MongoDB [12] was installed by downloading the ARM package and compiling the software on the NFS drive. Once installed, the database was easy to launch and a test database was defined, minimal datasets were loaded, and basic queries were executed against the database. The database is now available on the cluster for general R&D use. It seems to respond very well from a performance standpoint with the limited testing completed thus far. It also demonstrates the versatility of the platform by allowing for this non-traditional database software to be supported.

## IX. OBSERVATIONS, LIMITATIONS, AND FUTURES

### A. Relative Costs of the Octa Cluster

In comparing cost per performance of the Octa Board versus other computing environments we can see in Figure 9 below that the Octa Cluster provides a projected 399.6 GFLOPs at $5.01 per GFLOP. The comparative environments considered including the Raspberry Pi, a standard laptop, an enterprise level VMWare server intance, and a dedicated large scale physical Unix server have much lower computed performance levels and significantly higher costs. The original target system for replacement demonstration was a development machine which runs on an IBM RS/6000 P570 (server name REP-D1) running AIX and costs $875 per GFLOP per year. However this is an inaccurate comparison as the overall cost has to be incurred each year whereas with the cluster it is a one-time investment. We had difficulty demonstrating these exact computing capabilities but they provide a useful comparison metric.

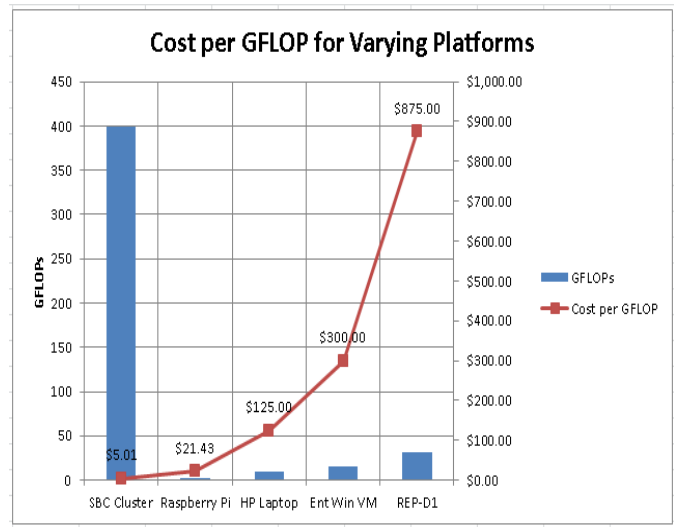

*Figure 9 – Cost vs GFLOPs on Octa-Cluster and several other computing platforms*

### B. Limitations of This Environment

The SBC based Octa-Cluster has a number of known limitations. These include the following:

- **Software Availability**: Some enterprise or standard software is not available for the ARM processor. We discovered this when looking at the possibility of running our Oracle database on the Octa-Cluster for development purposes. While the potential for costs savings was immense the reality is that Oracle does not offer a build of its Enterprise database for this platform. There are other options such as Berkely DB or MySQL and we may port our AIX based data files to this database but that would then require ongoing conversions back and forth adding time and effort. We continue to explore this problem for the right blend, for example, carrying out early prototyping on the cluster and the converting one time only over to our standard platforms.

- **OS Compatibility**: As an extension of the above problem, compatibility between development on Linux vs AIX or even Windows will continue to be an issue. By doing early application development under Linux this may cause later issues in compatibility on AIX environments. There could be a benefit here also as it may be possible to spot areas to work on to generalize the code base for future porting to Linux.
- **Support:** Naturally the Octa-Cluster is a fully custom environment and does not fall under any support agreement we have with our hosting vendor. In theory we could develop a custom agreement but that would be costly. As a result the device continues to require local administration. This means that 24x7 support will not be available by default. However, high levels of support can still be expected as the R&D team is more or less available and can work on the cluster when needed. Furthermore, there was essentially no support provided by the vendor especially when we posted question to the developer blog during our initial boot up challenges. Overall, few of the discussion threads really helped us very much.
- **Skillsets:** Within CT some of the required Linux administration, storage technology knowledge, or network engineering skills are not broadly available as these are provided by our hosting vendor who is not supporting this technology. Some of these skills were needed in differing amounts to build out the POC. Some creativity and learning was required to make this POC work. In most cases we helped each other and helped ourselves. Google searches for information became a very standard practice to getting to the solution of a problem.
- **Lack of redundancy:** It was unclear at the outset how hardened or reliable the device would be. Eventually it may be necessary to set up multiple clusters to provide adequate reliability and availability. In the case of the initial cluster, it has been running continuously for 6 months now without any forced reboots or failures. The environment does not have a RAID array. This would be a good investment in data protection for the future. Also, the cluster is running on conditioned power but without a UPS. That is another area for potential improvement.
- **Environmental controls:** The POC is physically located within a standard office space but on a dedicated lab desk. By placing the cluster in an office setting this obviously subjects the device to variations in temperature and other conditions. It might be possible to install the cluster in our local data center room with proper cooling and environmental controls but this complicates the team's access to the device. Up to now the physical environment has not seemed to provide any issue.
- **Hidden support costs:** Maintaining the POC obviously requires time from our staff. The administrative work is typically something our hosting vendor would do for our typical environments allowing CT staff to focus on development work and support work. It was expected that once the SBC cluster is set up and running there would be little significant administration required. Since the idea is to eventually give database privileges to the developers on this device it is assumed the support workload would actually go down. Nevertheless, this needs to be factored in with any such arrangement.

*C. Alternative Approaches*

Exploring SBCs are not the only possible approach to meeting the general needs of the problem statement presented at the beginning of this paper. There are at least several other approaches which we have discussed internally and may be pursued including:

1. Cloud Services – by placing some development environment resources in the facilities provided by alternate cloud vendors (outside our current private cloud) some cost savings could be achieved and new scalability options obtained. There is active work being done in researching these options. One limitation around this is the security requirements placed on us by some of our customers to maintain data in strictly defined environments.
2. Modified Data Sets – One reason the CT computing environments can be inflexible is due to overlapping demand for dedicated testing with specific data sets. Instead of restoring and masking 100% of large production data to development for testing it could be possible to restore a small subset and run development databases on any laptop. This would only require extraction scripts to be developed and thus the SBC environments would be less necessary.
3. Data Generation – In the same manner, instead of doing any data restoration, development and QA could generate test data which is infrequently done today. In this case, once again, the development or test databases could be run on nearly any computer whether standard or SBC. The scale of the data would be controlled by the developer.
4. Low Cost Servers – There are many low cost computer boards with large memory capacity available. If small form factor is not an issue then building custom low cost servers which are the size of typical PCs might be a possibility as well. The advantage of the SBC is not just the low cost but the small footprint and the possibility to cluster many of these devices together on a table top quickly as has been proven in this report.

*D. Future Possibilities*

SBCs and SBC centric clusters represents a non-linear jump in computing power of approximately 10x in its first generation. This technology outpaces current chip designs at the price and energy point provided. In one or two more generations this technology could be even more powerful and cost attractive. We believe it is beneficial to be on this curve early.

The SBC approach offers very low cost, high performance, low foot print computing. This changes the assumed model that bigger is better and even challenges the cloud computing model or the VM model. Why rent computing cycles at $100 a month when you can by more than you ever need for $100 as a onetime expenditure? At that cost you can replace most of your server infrastructure every year or two and increase performance perhaps by a factor of 3 each time. While our experience has been that the Octa Boards have proved reliable, even if they were to fail the replacement cost is so low it changes the economics of what might be considered for in house computing. Alternatively, these types of platforms may find their way into commercial Cloud datacenters and further drive down computing consumption costs.

If you need high end computing power simply scale out via clustering SBCs. This has already been proven to work with the Raspberry, the Octa Board, and other devices. In concept CT might build multiple development and test environments for extremely low cost for multiple applications thereby breaking the traditional environment bottleneck issues which we often encounter during peak project periods due to the lack of availability of computing resources. The use of such low power consumption devices may have very significant effects on costs and sustainability. Our cluster runs with 8 compact 5 volt power supplies and uses only one fan in order to provide air flow. Cooling is barely required as components on the Octa Boards in general do not heat up over time or dissipate any significant heat. The reduced cost around power and compact cooling systems of the future could provide dramatic benefits to a world facing energy constraints and $CO_2$ emission restrictions.

Some current generation SBCs might be limited in some areas such as RAM, I/O or other particulars. It is predictable that OEMs, 3rd parties, or open source providers will provide supplemental technologies to improve these areas. This is a young and dynamic field. Innovation will be rife and frequent. We will be looking to the next round of products and price for performance enhancements.

## X. CONCLUSIONS

Our journey to implement a High Performance Computing environment using a Single Board Computing Beowulf cluster has been mostly successful. We were able to build an initial Raspberry Pi prototype to learn about the capabilities of SBCs and then research available products to find a more powerful device in order to create a SBC HPC cluster. In creating the cluster built on InSignal's Octa Board and using an open source MPI library we were successful in running all 64 processing cores simultaneously putting a total of 96 GHz to use. We were also successful in running Elastic Search and several other software packages including ownCloud and MongoDB. We found the environment to be highly versatile and extensible with very straightforward administration requirements.

Where we ran into some difficulties was in finding business applications for the cluster. We thought we could port one of our business application databases to the cluster but as yet we have not done so. Also, while we were able to install Elastic Search we were unable to achieve usable results for the tool yet. Looking to the future we do plan to continue refining the cluster environment and attempt to demonstrate a clear business application to the cluster aside from its R&D function. In the end we do feel that for just a bit over our $2,000 budget we have built a highly capable computing platform which we can apply to many tasks in a flexible manner which fulfills the mission of the POC outlined at the outset of this paper. We also learned a lot at each step, enjoyed working on the platform, and have plenty of ideas of future projects stimulated by this work.

## XI. ACKNOWLEDGEMENTS

The creation, deployment, and operation of the Octa-Cluster was first inspired by the demonstration of a wide array of Raspberry Pi applications at the 12th Annual CUNY IT Conference presented by members of the CUNY Library staff [1]. Professor Rich Dragan of CUNY extended an invitation to this event. Further, the various published articles around the application of the Raspberry Pi as "supercomputers" and shared on the web helped get us started. Most importantly our CTO David Gardner was quick to realize the potential of what we were suggesting and agreed to support and fund the experiment.

## XIII. AUTHOR CONTACT

James Cusick, Director IT, Wolters Kluwer, New York, NY, j.cusick@computer.org.

Miller, William, Project Manager, Wolters Kluwer, New York, NY, william.miller@wolterskluwer.com.

Laurita, Nicholas Sr. Systems Engineer, Wolters Kluwer, New York, NY, nicholas.laurita@wolterskluwer.com.

Pitt, Tasha, Support Analyst, Wolters Kluwer, New York, NY, tasha.pitt@wolterskluwer.com.